\documentclass[aps,pre,twocolumn,showpacs,howkeys,preprintnumbers,amsmath,amssymb]{revtex4}

\usepackage{graphicx}
\usepackage{dcolumn}
\usepackage{bm}
\usepackage{amsmath}
\usepackage{amsfonts}
\usepackage{amssymb}
\usepackage[english]{babel}

\begin{document}
\title{Monte Carlo Simulations of the site-diluted 3$d$ XY model
with superexchange interaction: application to
Fe[Se$_2$CN(C$_2$H$_5$)$_2$]$_2$Cl -- Zn[S$_2$CN(C$_2$H$_5$)$_2$]$_2$ diluted
magnets}

\author{J. B. Santos-Filho$^{1}$ and J. A. Plascak$^{2,3,4}$ }
\address{$^1$ Instituto Federal de Ci\^encias e Tecnologia de Sergipe 
          \\  49100-000  S\~ao Crist\'ov\~ao, SE, Brazil}
\affiliation{$^2$ Universidade Federal da Para\'iba, Centro de Ci\^encias Exatas
e da Natureza - Campus I, Departamento de F\'isica - CCEN
Cidade Universit\'aria
58051-970 - Jo\~ao Pessoa, PB - Brazil}
\address{$^3$Departamento de F\'{\i}sica - ICEx, Universidade Federal de Minas
Gerais 
\\Caixa Postal 702, 30123-970 Belo Horizonte, MG, Brazil}

\address{$^4$Center for Simulational Physics, University of Georgia, Athens, GA
30602, USA}

\begin{abstract}
A simple site-diluted classical XY model is proposed to study the magnetic
properties of the ferromagnetic polycrystalline
Fe[Se$_2$CN(C$_2$H$_5$)$_2$]$_2$Cl diluted with diamagnetic
Zn[S$_2$CN(C$_2$H$_5$)$_2$]$_2$.  An extra superexchange interaction is assumed
between next-nearest-neighbors on
a simple cubic lattice, which are induced by a diluted ion. 
The critical properties are obtained by Monte Carlo simulations using a hybrid
algorithm, single histograms procedures and finite-size scaling techniques.
Quite good fits to the experimental results of the ordering temperature are
obtained, namely its much less rapidly decreasing with dilution than
predicted by the standard diluted 3$d$ XY model, and the change
in curvature around $86\%$ of the magnetic Fe[Se$_2$CN(C$_2$H$_5$)$_2$]$_2$Cl
compound.


\end{abstract}

\pacs{75.10.Hk, 75.30.Kz, 75.30.Hx, 75.40.Cx, 75.40.Mg}
\maketitle

The critical behavior of  disordered magnetic systems has been the subject of a
great amount of investigations during the last few decades both theoretically
and experimentally(see, for instance, \cite{stinch,liu} and references therein).
On the theoretical point of view, one of the extensively studied models is the
site diluted XY model. The XY model was originally introduced by Matsubara and
Matsuda \cite{Matsubara} to describe the behavior of liquid helium
\cite{santosfilho2}, but has also been quite suitable to describe the critical
behavior of some anisotropic insulating antiferromagnets
\cite{Betts,Betts2}. The three-dimensional (3$d$) version of the classical XY
model with site dilution also showed a
good agreement with some experimental results of some physical realizations,
such as the antiferromagnets Co$_{1-x}$Zn$_x$(C$_5$H$_5$NO)$_6$(ClO$_4$)$_2$
\cite{Algra} and [Co$_p$Zn$1_p$(C$_5$H$_5$NO)](NO$_3$)$_2$ \cite{Burriel}.
Recently, however, DeFotis et al. \cite{DeFotis} presented the phase diagram for
the pentacoordinate iron(III) molecular ferromagnet, in  which the simple site
diluted XY model predictions are in complete disagreement with the experimental
data. 

	The pentacoordinate iron(III) molecular ferromagnet
Fe[Se$_2$CN(C$_2$H$_5$)$_2$]$_2$Cl is, to date, the only known material to
exhibit three-dimensional XY ferromagnetic  behavior \cite{DeFotis2}.
Exchange interactions occur via intermolecular Se$-$Se contacts, the selenium
atoms being covalently bonded to the iron, leading to a ferromagnetic ordering
near $3.4$ K . The magnetic lattice is a simple cubic one and this random
compound, when diluted with non-magnetic
Zn[S$_2$CN(C$_2$H$_5$)$_2$]$_2$, presents an initial critical temperature slope
$s= \left[(dTc/dp)/Tc\right]_{p=1}=0.24(2)$ \cite{DeFotis}, where $p$ is the
concentration of the magnetic material Fe[Se$_2$CN(C$_2$H$_5$)$_2$]$_2$Cl (the
concentration of the non-magnetic material Zn[S$_2$CN(C$_2$H$_5$)$_2$]$_2$ is
$q=1-p$). This slope is well below the theoretical predictions for the $3d$
simple cubic XY
model with site dilution, whose value is $s=1.0965(39)$ \cite{santosfilho},
and even below the slope for the Ising and Heisenberg models. In addition,
there is also an inflection point in the critical temperature as a function of
the concentration $p$, which can not be accounted for by these simple diluted
models either. So, these striking
behavior seem not to be a model question, but some other microscopic feature
instead. 

In ref. \cite{DeFotis} the authors have indeed commented on a possible
superexchange pathway (although they claimed it seems not be so effective) between
Fe atoms which might be responsible for the weak decline in the critical
temperature with dilution. We will here pursue exactly in this direction and
show that a superexchange interaction can in fact account for this small decline
of the transition temperature with magnetic site concentration, and even for the
inflection point, with a good agreement with the experimental data.

So, in this work, without taking into
account the complex lattice structure of the pentacoordinate iron(III) molecular
ferromagnet, we propose an extended version of the simple site-diluted XY model,
defined on a simple cubic lattice, in the same lines as the works on Fe-Al
\cite{pla1, pla2} as well as on Fe-Al-Mn systems \cite{Alcazar}. In order to
study the phase diagram of this disordered system, we further assume herein that
the dopant, or the non-magnetic ion, can induce a superexchange like
ferromagnetic interaction between its
nearest-neighbor spins.

Thus, the system under investigation is given by a quenched site-diluted XY
model Hamiltonian that can be written as 
\begin{eqnarray}\label{hamil}
\begin{split}
\mathcal{H} =& -J_1\sum_{\langle i,j\rangle}\epsilon_i \epsilon_j
[S_{i}^{x}S_{j}^{x}+S_{i}^{y}S_{j}^{y}] \\& -J_2\sum_{\langle\langle
i,j\rangle\rangle}\epsilon_i \epsilon_j [S_{i}^{x}S_{j}^{x}+S_{i}^{y}S_{j}^{y}]
~,
\end{split}
\end{eqnarray}
where the first sum is over nearest-neighbors (NN) $\langle i,j\rangle$ spins, 
the second sum over next-nearest-neighbors (NNN)
$\langle\langle i,j\rangle\rangle$ spins, $\vec{S_i}$ represents a
three-dimensional classical spin $\vec{S_i}=(S^x_i,S^y_i,S^z_i)$, where
$S_i^\alpha$ are the $\alpha=x,~y,$ and $z$ cartesian components of $\vec{S_i}$
with $S_i^2=(S^x_i)^2+(S^y_i)^2+(S^z_i)^2=1$. $J_1>0$ is the NN
and $J_2>0$ is the NNN ferromagnetic interactions,
respectively. While the first sum in
Eq. (\ref{hamil}) runs over all nearest-neighbor pairs, the second sum runs only
over the next-nearest-neighbor pairs having a dopant as a common
nearest-neighbor, i.e., for situations in the lattice where a non-magnetic site
$k$ has the corresponding sites $i$ and $j$ as NN. In this
sense, in a simple cubic lattice, which is the case for the pentacoordinate
iron(III) molecular ferromagnet, depending on its concentration, a non-magnetic
molecule can break up to six interactions $J_1$ and, on the other hand, can
generate up to twelve interactions $J_2$. 

In Eq.(\ref{hamil}),  $\epsilon_i$ are quenched, uncorrelated random variables,
representing the existence of two kinds of particles in the system, namely the
magnetic ones with $\epsilon_i=1$, and non-magnetic
ones with $\epsilon_i=0$. The variable $\epsilon_i$ is chosen according to the
probability distribution
\begin{equation}
P(\epsilon_i) = p\delta(\epsilon_i-1) + (1-p)\delta(\epsilon_i),
\label{pro}
\end{equation}
where $p$ is the concentration of magnetic sites, such that $p = 1$ corresponds
to the pure case.

The above Hamiltonian has been studied through Monte Carlo simulations. First,
we prepare a diluted lattice where a given configuration of diluted sites
$\{\epsilon \}$ refers
to a single sample. For every thermodynamic observable $Q$, we first calculate
the thermal average $\langle Q_{\{\epsilon\}} \rangle$ for a given sample
$\{\epsilon \}$ and the results for different samples are later averaged as
$\left[ {\left\langle {Q_{\left\{ \epsilon \right\}} } \right\rangle }
\right]_{{\text{av}}}$. In order to get the critical properties of the present
model, for each sample of a given site-disorder configuration $p$, we used a
hybrid Monte Carlo algorithm consisting of one single spin Metropolis algorithm,
and one overrelaxation updates of the spins at constant configurational energy
\cite{creutz, pawig}. This hybrid Monte Carlo method~\cite{prb60} has 
been implemented, and has been shown to reduce correlations between successive
configurations in the simulation \citep{santosfilho}. Close to the  transition
temperature we have also resorted to single histogram techniques to get the
corresponding thermodynamic quantities. We have first computed the
in-plane-magnetization, magnetic susceptibility, and the Binder cumulant given,
respectively, by
\begin{eqnarray}
m_{xy}=\frac{1}{L^{2}}\sum_{i=1}^{L^{3}}[(S_{i}^{x})^{2}&+&(S_{i}^{y})^{2}],\\
\label{mag}
\chi_{} = L^{2} \frac{\langle m_{xy}^{2}\rangle -\langle m_{xy}\rangle
^{2}}{T}&,&~~u_{4}=1-\frac{\langle m_{x}^{4}\rangle}{3\langle m_{x}^{2}\rangle
^2},
\label{U1}
\end{eqnarray}
where $L$ is the linear size of the cubic lattice studied, $T$ is the
temperature given in units of $J/k_B$, $k_B$ being the Boltzmann constant, and
for the cumulant we have $m_\alpha^m = 
\left({\frac{1}{L^3}}\sum_{i=1}^{L^3}{S_i^\alpha}\right)^m$, with $\alpha=x,y,$
or $z$. We have also considered $J_1=1$, in such a way that $J_2$ is measured in
units of $J_1$. Although the natural, or standard, choice of the Binder cumulant
in the
context of an XY type transition would include both the $x$- and $y$-components
of the magnetization \cite{massimo}, it has been previously shown the $x$
component to be the most suitable one for computing $u_4$ and getting the
criticality of the model \cite{prb60}.

In addition, to reach the final results, for each dilution, temperature, and
lattice size, the MC estimates 
$\langle Q_{\{\epsilon\}} \rangle$ of thermodynamic quantities, for a given random distribution 
$\{\epsilon\}$ of diluted sites were averaged over different disorder
realizations as
\begin{equation}
\left[ {\left\langle {Q_{\left\{ \epsilon  \right\}} } \right\rangle } \right]_{{\text{av}}}  = \frac{1}
{{\# \left\{ \epsilon  \right\}}}\sum\limits_{\left\{ \epsilon  \right\}} {\left\langle {Q_{\left\{ \epsilon  
\right\}} } \right\rangle } ,
\label{av}
\end{equation}
with $\#\{\epsilon\}$ the number of total realizations considered. 

Now, regarding the simulational numbers, for every sample the runs comprised 
$ 10^3$ MCS per spin for equilibration and the measurements were made
over $5 \times 10^4$ MCS. The lattice sizes raged from $L=10$, $20$, $30$, $40$,
the values being chosen so that $p \times L^3$ gives an integer number. We have
used in the present paper $100$ samples for all settings. As discussed in ref.
\cite{santosfilho}, such simulations could give a good account of the model
with only NN interactions, even concerning the critical
exponents. We believe the same happens for this more general model, mainly  for
the critical temperature, which is our main purpose for the experimental data
we are comparing to.

The critical temperature  $T_c$  was obtained from the peak of susceptibility
and cumulant crossing for different values of $L$. We have used the correlation
length critical exponents of the XY universality class since it is independent
of the value of the dilution \cite{santosfilho}. The quality of the results are
the same as those obtained for the pure model and extensively discussed in ref.
\cite{santosfilho}. For this reason, we will only present herein the
corresponding results for the critical temperature.

\begin{figure}[htb]
\includegraphics[width = 7.4cm]{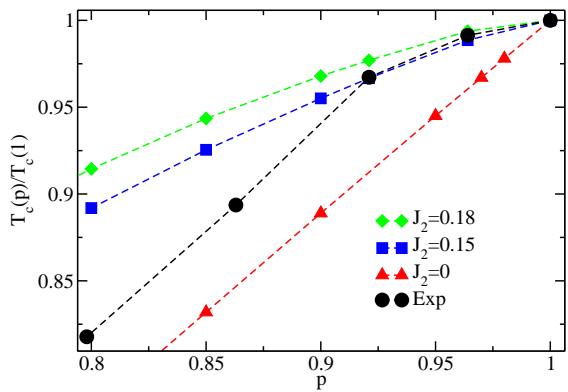}
\caption{(Color online) Reduced critical temperature $T_c(p)/T_c(1)$ as a
function of the mangnetic sites concentration $p$, for several values of the
next-nearest-neighbor interaction $J_2$ ($J_1=1$). The experimental results,
represented by the circles, were taken from Ref.\cite{DeFotis}. The other
symbols are simulation results. The dasheds lines are just guide to the eyes.
The simulational errors are smaller than the symbol sizes.}
\label{fig1}
\end{figure}

Fig. \ref{fig1} exhibits, for several values of the
NNN interaction $J_2$, the reduced transition temperature
$T_c(p)/T_c(1)$ as a function of $p$, where $T_c(p)$ is the critical temperature
for the concentration $p$ and $T_c(1)$ is the transition temperature of the pure
system. In this case, as just a matter of comparison, we have considered
ordinary next-nearest-neighbor interactions $J_2$, in the sense that no
superexchange character has been implemented. It means that second-neighbors
interact, independently of their neighborhood. One can clearly see that, as the
next-nearest-neighbor interactions increase, the slope close to $p=1$ decreases.
Nevertheless, no value of $J_2$ can give a satisfactory behavior of the
transition temperature in the wider range of concentrations analysed
experimentally. This means that the dilution alone is not responsible for the
experimental phase diagram.

The situation is indeed completely different when we further assume that the
NNN interaction $J_2$ is induced by a diluted site, or a
dopant, in a kind of superexchange interaction. Recall that, in this case, only
impurities whose nearest neighbors are magnetic sites enable the arising of a
superexchange interaction between NNN magnetic sites. 
In addition, another effect should be included in the simulations. In Fig.
\ref{fig0}(a) one has a square lattice view of a non-magnetic dopant where four
NNN interactions are induced. However, when one has a cluster of two dopants, as
in Fig. \ref{fig0}(b), they tend to get closer, since the unit cell of the
selenium material is 11.3\% larger due to the greater size of Se versus S
\cite{DeFotis}. So, the distance from the dopant tends to increase in this case,
and one expects a smaller superexchange interaction $J_2^\prime$. The same will
happen for a $J_2^{\prime\prime}$ when three dopants are closer, and so on. When
we consider $J_2=J_2^\prime=J_2^{\prime\prime}=...$, we recover the ordinary NNN
interaction model results shown in Fig. \ref{fig1}.  Nevertheless, by
assuming that only $J_2$ is sufficiently high (i.e.
$J_2^\prime=J_2^{\prime\prime}=...=0$) we arrive at a global phase diagram 
depicted in Fig. \ref{fig2}. In this figure we have the results for $J_2=0.18$
and $J_2=0$ as well. One can notice now a good
agreement with the experimental results. The theoretical critical line of the
$J_2=0$ model goes to the limit of the percolation threshold of the simple cubic
lattice (we have not done simulations for $p<0.4$, because in this case the
transition temperature is very small). Moreover, for larger non-magnetic
concentrations, the results from the model considering next-nearest-neighbor
interactions approach those of the pure model. This is understandable, since the
higher the dilution, the more difficult it becomes to create new superexchange
interactions, and the system ends up with only NN interactions (a
fact that does not happen in the model with ordinary NNN
couplings). 

\begin{figure}[!t]
   \hspace{0.8cm}
 {\scriptsize\textbf{(a)}}\\
   \includegraphics[height=4.5cm]{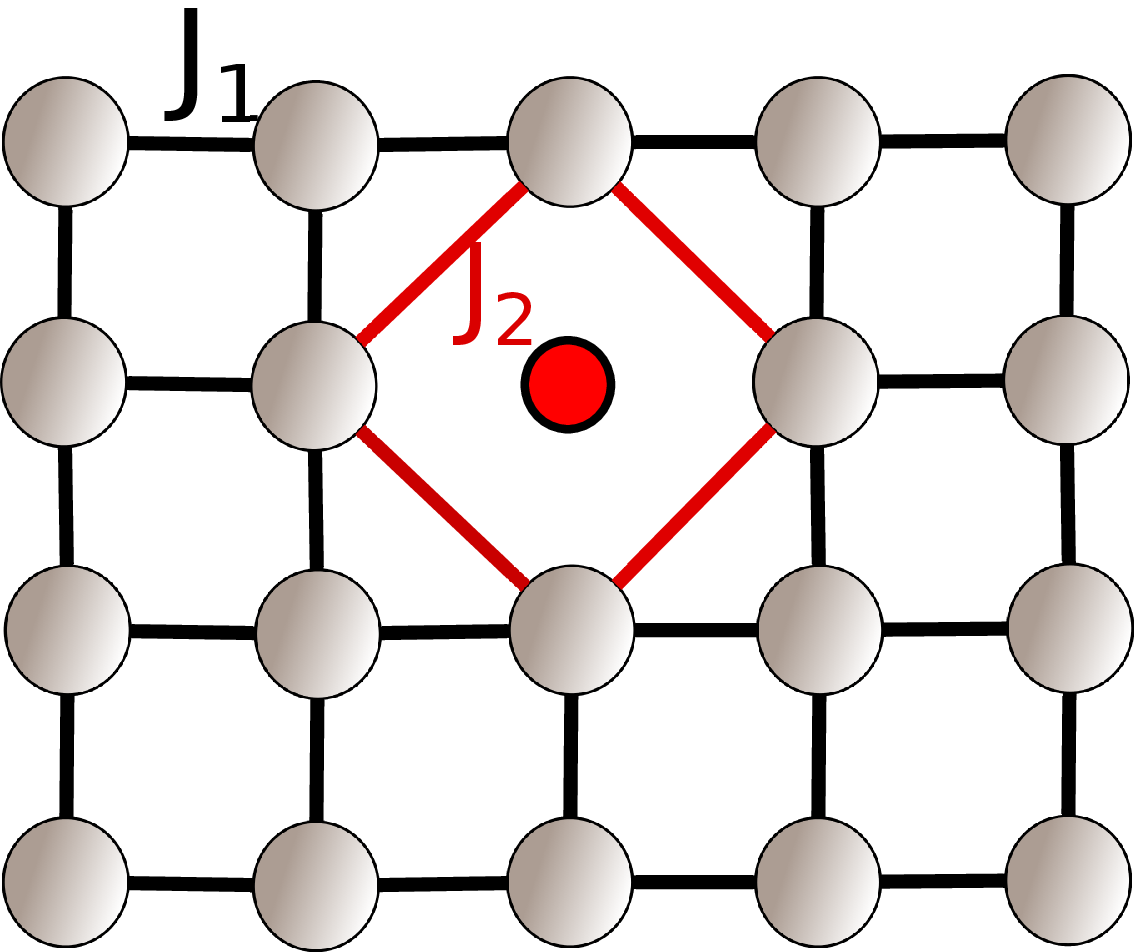}
\end{figure}
%
\begin{figure}[!t]
   \centering
   \hspace{0.8cm}
 {\scriptsize\textbf{(b)}}\\
   \includegraphics[height=4.5cm]{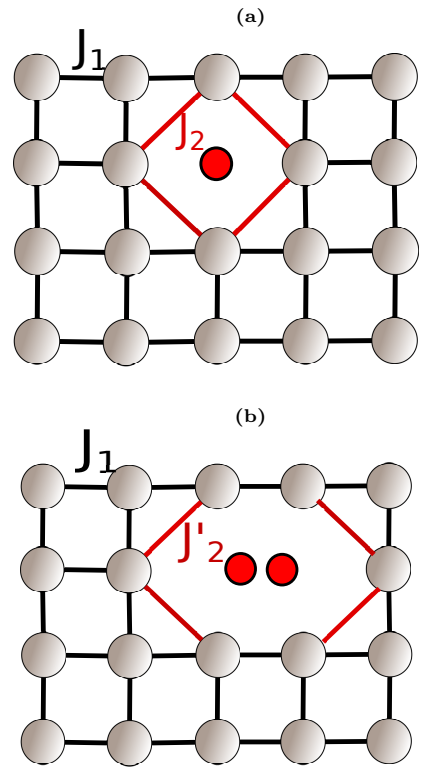}
\caption{ (Color online) A square lattice sketch of the induced
superexchange interaction $J_2$ by the dopant (isolated circles). In (a), one
dopant generates the NNN interaction $J_2$. In (b), a cluster of two dopants
(on an exaggerated scale). They are now farther from the magnetic ions and one
has a NNN $J_2^\prime<J_2$.}
\label{fig0}
\end{figure}

\begin{figure}[htb]
\includegraphics[width = 7.4cm]{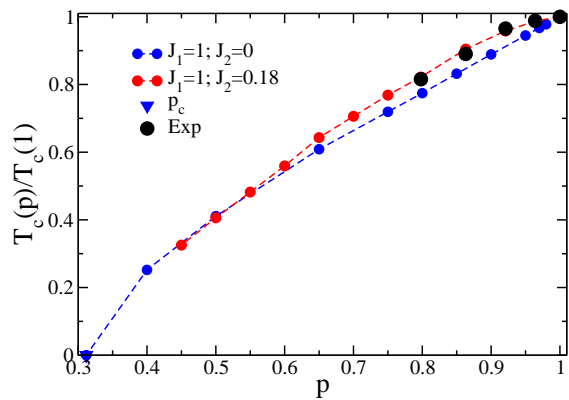}
\caption{(Color online) The same as Fig. \ref{fig1}, for $J_2=0$ and $J_2=0.18$,
assuming that the superexchange interaction $J_2$ is induced by the presence of
a diluted site. $p_c$ is the percolation threshold for the simple cubic
lattice.}
\label{fig2}
\end{figure}

\begin{figure}[htb]
\includegraphics[width = 7.4cm]{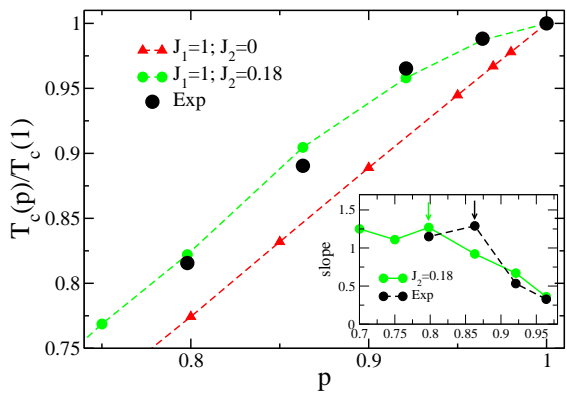}
\caption{(Color online) A closer view of the phase diagram from Fig. \ref{fig2}
in the region near the pure system for $J_2=0$ and $J_2=0.18$. The inset shows
the estimate of the corresponding slopes as a function of $p$. The arrows
indicate the inflection point. Regarding the simulations, the errors are smaller
than the symbols sizes.}
\label{fig4}
\end{figure}

It is also interesting to look at the phase diagram closer to the pure model,
as is shown in Fig. \ref{fig4}, where the discrepancy from the simple site
diluted model and the one considering superexchange next-nearest-neighbor
interactions is clearer still.  In addition to the quite good fits to the
experimental data, one can also note, from the inset in Fig. \ref{fig4}, that
the present model also exhibits an inflection point around $p=0.80(5)$, close to
$p=0.86$ that is experimentally observed. 

In summary, we have proposed a simple diluted XY model on a simple cubic
lattice where superexchange interactions are induced between
NNN via dilution sites. Not only the slow slope of the
critical temperature as a function of the concentration agrees well with the
experimental data, but also an inflection point of the critical curve is
present, and quite close to the corresponding experimental concentration. Of
course, this is only a theoretical assumption still needing a more
experimental evidence.

A little more words are, however, worthwhile in commenting the present
simulations. First, it is well known that MC simulations are not suitable in
adjusting theoretical parameters to experimental data, mainly due to
computational costs. In the present case things are better in the sense that we
have just one parameter. As one knows the critical temperature for the pure XY
three-dimensional model with reasonable accuracy, this serves to estimate
$J_1$. But, renormalizating the critical temperature data by its value for the
pure model, we can get rid of $J_1$ and just fit in fact the ratio $J_2/J_1$.
So, as we can further consider $J_1=1$ in our equations, all the simulations
have to be done for different values of the next-nearest-neighbor coupling
ratio. This makes the fitting problem an easier one. 

As a second remark, we recall that we have used FSS to get
$T_c(p)$, so the corresponding values of the critical temperature are indeed
very accurate. This means that any problems with the comparison to the
experimental data, if any, would be related to the quantum nature of the spin
state, since we have considered a classical spin model. However, as Fe ion can
have a spin-$3/2$, the use of a classical model can, in some sense, be suitable
for the present case.  

As a final comment, from the critical temperature of the pure model
$k_BT_c/J_1=1.5518$ \cite{santosfilho} one can estimate the nearest-neighbor
interaction $J_1=0.189$ meV, which should be compared to $12.8$ meV obtained for
the Fe-Mn-Al alloys \cite{pla1,pla2,Alcazar} (or to $10-50$ meV as is usual for
ferromagnetic systems). These two orders of magnitude smaller can be understood
since for the present system the ordering temperature is very low, near $3.4$ K,
instead of about $1000$ K, as is the case of the Fe based alloys. On the other
hand, the ratio $J_2/J_1=0.18$ is within of what is expected for a
second-neighbor interaction. Of course, more theoretical and experimental studies 
on these systems would be very welcome.

\acknowledgements{
The authors would like to thank financial support from CNPq, CAPES, and CNPq grant
402091/2012-4.}

 \end{document}